\documentstyle[12pt,amstex,righttag]{article}
\numberwithin{equation}{section}

\def\g{\overline{g}}
\def\phibar{\overline{\phi}}
\def\psibar{\overline{\psi}}
\def\Fbar{\overline{F}}
\def\reg{\theta(d\cdot i/n )}

\begin{document}
\begin{titlepage}
\begin{flushright}
	KEK-TH-475 (revised)	\\
	August  1996
%\\ {hep-th9604022}	
\end{flushright}
\vspace{5mm}
%%%%%%%%%%%%%%%%%%%%%%%%%%%%%%%%%%%%%%%%%%%%%%%%%%%%%%%%%%%%%%%%%%%%%%%%%%%%%%%%%
\begin{center}
{\LARGE
RESIDUES AND TOPOLOGICAL YANG-MILLS \\ \vspace{0.5cm}
THEORY IN TWO DIMENSIONS   }  \\
\vspace{2cm}
{\large {\sc Kenji Mohri}}
%\footnote{{\it e-mail address} mohri@@theory.kek.jp}
 \\
\vspace{1cm}
{\it 
National Laboratory for High Energy Physics (KEK) Ibaraki, JAPAN 305} \\ 
\vspace{0.5cm}
 mohri@@theory.kek.jp
\end{center}
\vspace{2cm}

%%%%%%%%%%%%%%%%%%%%%%%%%%%%%%%%%%%%%%%%%%%%%%%%%%%%%%%%%%%%%%%%%%%%%%%%%%%%%%%
\begin{abstract}
%ABSTRACT NONSENSE\\
A  residue formula which evaluates 
any correlation function  of topological $SU_n$ Yang-Mills theory 
with arbitrary magnetic flux insertion  in two dimensions  are obtained.
Deformations of the system  by two form operators 
are investigated  in some detail.
The method of the diagonalization
of a matrix valued field
turns out to be useful to compute various physical quantities.   
As an application  we find the operator that contracts a handle 
of a Riemann surface and a genus recursion relation.

\end{abstract}
\end{titlepage}
\baselineskip=0.8cm
%
%%%%%%%%%%%%%%%%%%%%%%%%%%%%%%%%%%%%%%%%%%%%%%%%%%%%%%%%%%%%%%%%%
\section{Introduction}
Two dimensional topological Yang-Mills theory 
is an example of topological field theories which is 
precisely the two dimensional analogue of 
the Donaldson theory \cite{Witten5}.\\
The physical observables of this theory 
in the favorable cases are identified with the cohomology classes of 
the moduli space ${\frak M}$
of the flat gauge fields on a Riemann surface.\\
The cohomology ring of  ${\frak M}$  \cite{Thadd,Bara,BeJoSaVa,Zagier2}
 recently draws  much attention
in connection  with the Floer cohomology group \cite{DoSa}. 
The correlation functions
of topological Yang-Mills theory,
which determine the cohomology ring,
  have been completely solved in the form of 
a multiple infinite sum  
 by Witten using the exact solution of  two dimensional
physical Yang-Mills theory  \cite{Rusakov,BlaTho,Witten3}
and the correspondence between  physical and topological Yang-Mills
theories \cite{Witten4}.\\
Apart from the unsolved problem of the cases of non compact gauge groups
which may be important to analyze topological $W$ gravities, 
there is still  much to do in this theory.\\
%%%%%%%%%%%%%%%%%%%
First of all  it seems  difficult to compute 
the explicit value of  a correlation
function  and to investigate the cohomology ring 
of topological Yang-Mills theory of gauge groups  other than $SU_2$ 
using the infinite sum  formula found in \cite{Witten4}.
Thus it would be  desirable to
find  another formula of correlation functions
which directly gives their values as rational numbers and is more suitable 
to study the cohomology ring.\\
Moreover the formula in \cite{Witten4}
for the correlators 
that contain general two form operators has remained  totally implicit
because we must  perform inversion of 
matrix field variables to evaluate it.  \\
%Thus the effective method of inversion are needed here.\\
%Third  investigation of topological Yang-Mills theory 
%with non compact gauge groups would be useful  
%to study  topological $W$ gravity.\\
%
In this paper we give a formula 
which expresses  any correlation function
of $SU_n$  theory with arbitrary 
magnetic flux  as a  residue evaluated
at the origin of the Cartan subalgebra
 generalizing the previous results  \cite{Thadd,Szenes}.\\
In the process we develop systematically 
the method of inversion of variables
using residues and the diagonalization of matrix valued field. \\
In addition to  its practical value, this residue formula 
 sheds light on  the structure of
the topological Yang-Mills  theory.
For example,  topological Yang-Mills theory may be regarded as a 
kind of matrix models 
 because the evaluation of correlation functions
by the  residue can be written as an  integration 
over eigenvalues of Hermite matrixes.\\
We also expect that the residue formula 
will be useful to consider systematically  the quantum deformation
of the cohomology ring \cite{BeJoSaVa},
 the coupling of topological Yang-Mills
theory to topological gravity and string interpretation of 
the large $N$ expansion
of topological Yang-Mills theory \cite{Moore}.\\
Jeffrey and Kirwan have studied the non-Abelian localization 
in \cite{JeffKir} and proved the residue formula for $SU_2$ \cite{JeffKir2}.\\ 
Blau and Thompson have studied  two dimensional topological Yang-Mills
theory as well as other closely related gauge theories 
based on the path integral approach.\\
They introduced the Abelianization, i.e. ,
 the diagonalization of the matrix valued
fields  in \cite{BlaTho2,BlaTho2.5},  which naturally reproduces 
 the infinite sum formula mentioned above.  
In \cite{BlaTho3}, the Abelianization was related to the localization 
of the gauge field path integral
to the reducible gauge fields using the equivariant  supersymmetry.
The relation between  this localization and the non-Abelian localization
to the Yang-Mills connections seems  still mysterious.
This paper is organized as follows. In Sect.2.1 
the magnetic flux  for gauge theory 
on a two dimensional surface is introduced.
Sect.2.2 is devoted to the standard construction of the BRST observables.
In Sect.3 we review the  result about physical Yang-Mills theory
and the infinite sum formula for correlation functions of topological 
Yang-Mills theory. In Sect4, we give the residue formula for 
the correlation functions with arbitrary magnetic flux
which contain  no two form operators  but  the symplectic form
$\omega$. 
We also describe some concrete examples of the correlation functions.
In Sect.5 we treat the cases in which 
 arbitrary two form operators are inserted in a correlator.
This  corresponds to considering  deformations 
of the original topological theory.
The most general residue formula for  the correlation functions
is obtained.
Finally in Sect.6, as an application of the residue formula  we find the 
observable that  restricts the path integral
to the subspace where the holonomy of any gauge field 
around a cycle is trivial and does nothing else.

\section{Topological Yang-Mills on a Riemann Surface}
\subsection{'t Hooft Magnetic Flux }
Here we describe the geometrical setting for 
 $SU_n$ topological Yang-Mills theory
on a genus $g$ Riemann surface $\Sigma_{g}$.
We introduce 't Hooft magnetic $\text{flux}^{\text{a}}$,
which is the terminology originally used 
in  four dimensional gauge theories \cite{Witten4,tHooft,VaWi},
as follows.\\
Pick a point $P$ on $\Sigma_{g}$
and put the boundary condition on gauge fields 
that the holonomy  around $P$ must be  $X^d$, where
%\begin{equation}
%\newcommand{\bg}{\family{crm}\size{20}{12pt}\selectfont}
%\newcommand{\bigzerol}{\smash{\hbox{\bg 0}}}
%\newcommand{\bigzerou}{\smash{\lower1.7ex\hbox{\bg 0}}}
\begin{equation}
%$$
X=\text{diag}
\left(e^{\frac{2\pi\sqrt{-1}}{n}},...,e^{\frac{2\pi\sqrt{-1}}{n}}%
\right)
%$$
\end{equation}
%M(n)=\begin{pmatrix}
%e^{\frac{2\pi \sqrt{-1}}{n}}& & & \bigzerou \\
% & e^{\frac{2\pi \sqrt{-1}}{n}} & & \\
%& & \ddots  & \\
%\bigzerol  & & & e^{\frac{2\pi \sqrt{-1}}{n}}\\
%\end{pmatrix}
%\end{equation}
%
 is a generator of the center ${\Bbb Z}_n$ of $SU_n$.\\
This is called the $SU_n$ gauge theory with $d$ units of  magnetic flux.
Note that the magnetic flux $d$ is  defined only modulo $n$ 
and the theory with $d$ units of magnetic flux   are related to  one 
with $(n-d)$ units of magnetic flux
 by the charge conjugation.
Modulo the degrees of freedom of 
the ghost zero modes associated with the residual
gauge symmetries
for the cases of $(n,d)\ne 1$, path integral of the topological theory
can be regarded as a integration over the moduli space of 
 flat gauge fields on $\Sigma_g-P$ 
with the prescribed holonomy around $P$, 
which we denote as  ${\frak M}_{g}(n,d) $.
${\frak M}_{g}(n,d)$ is also the absolute minima of 
the ordinary Yang-Mills action.\\
We list  some  properties of ${\frak M}_{g}(n,d)$ 
relevant to the later discussion;
\begin{itemize}
\item ${\frak M}_{g}(n,d)$ has the  
real dimension equal to the ghost number violation  $2(n^2-1)(g-1)$.
\item  ${\frak M}_{g}(n,d)\cong {\frak M}_{g}(n,n-d)$ 
due to the charge conjugation symmetry.
\item When $(n,d)=1$, ${\frak M}_{g}(n,d)$ is  smooth 
and has  no reducible gauge fields.
\end{itemize}
In the case $(n,d)\ne 1$, ${\frak M}_{g}(n,d)$  always 
has  reducible gauge fields, which make the analysis of the theory
  difficult.
In this paper  we define  correlation functions of 
 topological Yang-Mills theory as  
expansion coefficients
in coupling constant  of the partition function $Z(\epsilon)$ of 
 physical Yang-Mills theory \cite{Witten4}.   
This procedure indeed gives the correct answer  
for the case $(n,d)=1$.
For the other cases $(n,d)\ne 1$, in addition to 
a polynomial part, there also exist non-local terms in $\epsilon$
in the expansion of  $Z(\epsilon)$ due to the reducible flat gauge fields,
which makes the identification
of topological  correlation functions with
the polynomial part in $\epsilon$ of $Z(\epsilon)$
somewhat doubtful.
%
%ere is no guarantee that 
%the result given by this approach coincides with the one obtained 
%by the ``standard field theoretical method'' 
%and should be taken with  a grain of salt.
%In fact it was argued in \cite{Witten4} that
%for the cases  $(n,d)\ne 1$
%a correlation function  is  well-defined only when it  
% contains sufficiently many two form operators.
However  even in these cases the result obtained by this approach 
is consistent 
with the  Riemann-Roch-Verlinde formula, as we will see later,
with the reservation that 
 the existence of the Riemann-Roch formula itself on a moduli space 
with singularities is also presumed.
Thus   we will not worry about the singularities 
associated with reducible flat gauge fields in the cases $(n,d)\ne 1$ 
as long as  correlation functions are concerned.   
For the basic facts about  the Lagrangian and 
BRST symmetry  see \cite{Witten4,BlaTho2.5,Park}. 
%
%%%%%%%%%%%%%%%%%%%%
\subsection{Topological Observables}
Here we review the standard construction of the BRST observables of 
the topological theory \cite{Park}.
 Mathematical treatments of this subject for
 general $SU_n$ gauge group can be found  in \cite{AtBo,Earl}.
Let 
$(A_{i},\psi_{i},\phi)$
be the basic topological multiplet of the topological Yang-Mills theory, with
the BRST symmetry
$\delta A=\psi,\ \delta \psi=-D\phi,\ \ \delta\phi=0$.\\
It is useful to redefine
the fields as
$(\overline{\phi},\overline{\psi},\overline{F})=
(\frac{\sqrt{-1}}{2\pi}\phi,\frac{\sqrt{-1}}{2\pi}\psi,
\frac{\sqrt{-1}}{2\pi}F)$.\\
There are  two ways of component expansions of the matrix-valued
field $\phibar$.
The first one is the expansion with respect to
 an orthonormal basis of Lie algebra,
$
\phibar=\sum_{a}\phibar^{a}J_a.
$ \\
The second one, which turns out to be  more useful, is defined using 
a diagonalization 
\begin{equation}
\phibar=\text{diag}(z_1,...,z_n)=\sum_{i=1}^{n-1}x_iH_i,\ \ \ %
x_i= z_i-z_{i+1}=\langle \alpha_i,\phibar \rangle,
\end{equation}
where $\{ H_i \}$ and 
$\{ \alpha_i \}$ are the set of fundamental coweights and the simple roots
of $SU_n$ respectively, and the Weyl group action is simply the 
permutations of $\{ z_i \}$.\\
Now  the zero form operator of ghost number $2m$ 
 is defined by 
\begin{equation}
{\cal O}_{m}=\frac{1}{m!}\text{Tr}(\phibar^{m}).
\end{equation}
%%%%%%%%%%%%%%%%%%%%%%%%%%%%%%%%%%%%%%%%%%%%%%%%%%%%%%%%%
Next one form  operator of ghost number $(2m-1)$ is
\begin{equation}
 V_{m}(a)=\oint_{C_{a}}\frac{1}{m!}m\text{Tr}(\phibar^{(m-1)}\psibar), 
\end{equation}
%%%%%%%%%%%%%%%%%%%%%%%%%%%%%%%%%%%%%%%
where $ C_{a}, 1\le a \le 2g $ are  the 1-homology basis of $\Sigma_{g}$
such that $C_a\cdot C_{b+g}=-\delta_{ab}$.\\
Note that $\{V_{m}(a)\}$ transform among themselves 
under the mapping class group of $\Sigma_{g}$ 
\cite{Thadd,BeJoSaVa} and only the modular invariant combination 
of those;
\begin{equation}
\Xi_{lm}
%\int_{\Sigma}{\cal O}_{l}^{(1)}{\cal O}_{m}^{(1)}
=\sum_{a=1}^{g}(V_l(a)V_{m}(a+g)-V_{l}(a+g)V_{m}(a))
\end{equation}
are nonvanishing in the correlation functions.\\
Finally two form operator of ghost number $(2m-2)$ is 
similarly constructed as
\begin{equation}
{\cal O}_{m}^{(2)}=-\frac{1}{m!}\int_{\Sigma}\text{Tr}
(m\phibar^{(m-1)}\Fbar+\frac{1}{2}m(m-1)\phibar^{(m-2)}\psibar^{2} ).
\end{equation}
%
%
%%%%%%%%%%%%%%%%%%%%%%%%%%%%%%%%%%%%%%%%%%%%%%%%%%%%%%%%%%%%%%%%%
In particular, there always exist  observables
 associated with  the degree two  Casimir invariant;
${\cal O}_{2}$ and 
$\omega={\cal O}_{2}^{(2)}$ which is the standard symplectic two 
$\text{form}^{\text{b}}$  on ${\frak M}_{g}(n,d)$.\\
They  play the special role in physical/topological Yang-Mills 
correspondence \cite{Witten4}.\\
The correlation function of the form;
\begin{equation}
\left\langle 
\prod_{i}{\cal O}_{l_i}\prod_{j}V_{m_j}(a_j)\prod_{k}{\cal O}^{(2)}_{n_k},
\big[{\frak M}_{g}(n,d)\big] \right\rangle 
\end{equation}
is non-vanishing only if the observables inside  the correlator satisfy 
the ghost number selection rule,
\begin{equation}
\sum_{i}2l_i+\sum_{j}(2m_j-1)+\sum_k(2n_k-2)=2(n^2-1)(g-1).
\end{equation}
Hereafter we  will frequently use the notation $\g=(g-1)$.

%%%%%%%%%%%%%%%%%%%%%%%%%%%%%%%%%%%%%%%%%%%%%%%%%%%%%%%%%%%%%%%%%%%%%%%%%%%%%%%%
\newpage
%%%%%%%%%%%%
\section{Physical Yang-Mills Theory}
\subsection{Infinite Sum Formula}
In  \cite{Rusakov,BlaTho,Witten4},  a  multiple  infinite sum formula  
was obtained for the partition function of
 {\it  physical Yang-Mills} theory .\\
Physical Yang-Mills theory can be 
described  by the same field content as the basic BRST multiplet 
of the topological Yang-Mills
so that the generalized Lagrangian of the model with a polynomial 
$Q(x)=\epsilon \dfrac{x^2}{2!}+\sum_{m\geq 3}\delta_{m}\dfrac{x^m}{m!}$
%+\delta_3 \dfrac{x^3}{3!}+\delta_4 \dfrac{x^4}{4!}+\cdots$,  is 
with nilpotent $\{\delta_m\}$ is given by  
\begin{equation}
L=-\text{Tr}Q(\phibar)+\text{Tr}(\phibar \Fbar+
\frac{1}{2}\psibar \psibar).\label{PhysYangMills}
\end{equation}
Then the partition function is written as the following 
multiple infinite sum closely related to the multiple zeta value 
investigated in \cite{Zagier}   
%%%%%%%%%%%%%%%%%%%%%%%%%%%%%%%%%%%%%%%%%%%%%%%%%%%%%%%%
%      %
%                                                      %
%                                                      %
%%%%%%%%%%%%%%%%%%%%%%%%%%%%%%%%%%%%%%%%%%%%%%%%%%%%%%%%
%
%
\begin{equation}
Z(Q,\omega)=(-1)^{(n-1)d+|\Delta_{+}|\g}n^{g}
\sum_{l_1=1}^{\infty}\cdots \sum_{l_{n-1}=1}^{\infty}
\prod_{\alpha \in \Delta_{+}}\left<\alpha,\Phi \right>^{-2\g}
e^{\text{Tr}Q(\Phi)}
e^{-d\left<\lambda_{1},\Phi \right>},\label{zeta}
\end{equation}
%%%%%%
 where we set $\Phi=2\pi\sqrt{-1}\sum_{i=1}^{n-1}H_{i}l_i$ 
and $H_{i}$ is the $i$-th fundamental coweight .\\
%%%%%%%%%%%%%%%%%%%%%%%%%%%%%%%%%%%%%%%%%%%%%%%%%%%%%%%%%%%%%%%%%%%%%%%%%%
%%%%%%%%%%%%%%%%%%%%%%%%%%%%%%%%%%%%%%%%%%%%%%%%%%%%%%%%%%
%%%%%%%%%%%%%%%%%%%%%%%%%%%%%%%%%%%%%%%%%%%%%%%%%%%%%%%%%%
The normalization of the partition function %${}^{\text{c}}$ 
is determined so as to produce the corresponding 
topological correlation function by the expansion in the coupling constant
 $\epsilon$.
%%%%%%%%%%%%%
%%%%%%%%%%%%%%%%%%%%%%%%%%%%%%

%%%%%%%%%%%%%%%%%%%%%%%%%%%%%%%%%%%%%%%%%%%%%%%%%%%%%%%%
\subsection{Non-Abelian Localization }
Due to the {\it non-Abelian localization theorem} 
\cite{Witten4,JeffKir}, see also \cite{BlaTho3}, 
the path integral of physical Yang-Mills theory
can be localized around the solution of the equation of motion
which are the critical points of $S_{YM}$.
Noting  that the flat gauge fields are
the absolute minima which give the dominant contribution, it is seen that
 the partition function of the  physical Yang-Mills theory has  
 the following connection with the correlation function 
of the topological theory in the case of $(n,d)=1$,
\begin{equation}
Z(Q,\omega)=
\left\langle  e^{\omega}e^{\text{Tr}Q(\phibar)},     
    \big[{\frak M}_{g}(n,d)\big] \right\rangle
+\left\{\text{contributions of non-flat solutions}\right\}.\label{local}
\end{equation}
The first term  of the right hand side of (\ref{local})
is a polynomial in $\epsilon$ and represents  the  
corresponding correlation function of 
 topological Yang-Mills theory, while the  second term 
means  the contributions to the path integral 
from the solutions of the Yang-Mills equation
with non-zero action \cite{AtBo}
which has  the $\epsilon$-dependence 
$\simeq \exp(-\frac{ S_{YM}}{(2\pi)^2\epsilon})$.\\
More precisely it was shown \cite{Witten4} that the infinite sum of the form
\begin{equation}
\sum_{l_1=1}^{\infty}\cdots \sum_{l_{n-1}=1}^{\infty}
\prod_{\alpha \in \Delta_{+}}\left<\alpha,\Phi \right>^{-2\g}
e^{\epsilon\text{Tr}\Phi^2/2}\text{Tr}B(\Phi)
e^{-d\left<\lambda_{1},\Phi \right>}
\end{equation}
vanishes exponentially for $\epsilon \rightarrow 0$
if $(n,d)=1$ and the ghost number of $B$ is greater than $4\g |\Delta_{+}|$.
When $(n,d)\ne 1 $, reducible flat gauge fields 
give additional terms which are non-local in the coupling constant.
Even in these cases the polynomial part could be given the 
interpretation as  topological correlation functions.\\
Thus  we obtain the infinite sum formula
for correlation functions of topological Yang-Mills theory
which is somewhat conjectural for $(n,d)\ne 1$ cases;
\begin{align}
&\left\langle  e^{\omega}\text{Tr}B(\phibar),     
    \big[{\frak M}_{g}(n,d)\big] \right\rangle\nonumber \\
&=(-1)^{(n-1)d+|\Delta_{+}|\g}n^{g}
\sum_{l_1=1}^{\infty}\cdots \sum_{l_{n-1}=1}^{\infty}
\prod_{\alpha \in \Delta_{+}}\left<\alpha,\Phi \right>^{-2\g}
\text{Tr}B(\Phi)
e^{-d\left<\lambda_{1},\Phi \right>}.\label{sumformula}
\end{align}
The above formula  is well-defined  only when  the 
right hand side  converges.
This in particular implies that
 the ghost number of $B$ should be less than
$4\g |\Delta_{+}|$ in (\ref{sumformula}).
%%%%%%%%%%%%%%%%%%

%%%%%%%%%%%%%%%%%%%%%%%%%%%%%%%%%%%%%%%%%%%%%%%%%%%%%%%%%%%%%%%%%%%%%%%%%%%%%
%\newpage
%%%%%%%%%%%%%%%%%%%%%%%%%%%%%%%%%%%%%%%%%%%%%%%%%%%%%%%%%%%%%%%%%%%%%%%%%%%%%%%%
\section{Residue Formula}
%\newpage
%%%%%%%%%%%%%%%%%%%%%%
%%%%%%%%
\subsection{Residue Form with  Magnetic Flux}
Here we propose the residue formula for correlation functions 
of topological Yang-Mills theory with arbitrary magnetic flux.
For the case of $(n,d)=(2,1)$ and the case of $(n,0)$
were treated by  Thaddeus \cite{Thadd}
and Szenes \cite{Szenes} respectively.\\ 
To present the formula we introduce the following symbol;
\begin{equation}
\theta\left(\frac{d\cdot i}{n}\right)=\frac{d\cdot i}{n}
-\left[\frac{d\cdot i}{n} \right].
\end{equation}
%%%%%%%%%%%%%%%%%%%%%%%%%%%%%%%%%%%%%%%%%%%%%%%%%%%%%%%%%%%%%%%%%%%%%%
Now  define the multi-variable residue form $\Omega^{(k)}_g(n,d)$ 
for $SU_n$ theory with $d$ units of magnetic flux by 
\begin{align}
\Omega_{g}^{(k)}(n,d) =&
(-1)^{(n-1)(d-1)+|\Delta_{+}|\g}
\left(nk^{n-1}\right)^{\g}
\prod_{i=1}^{n-1} 
%\underset{x_i=0}{\text{Res}}
\dfrac{dx_{i}}{x_{i}}
\dfrac{kx_i}{(e^{kx_{i}}-1)}\nonumber \\
&\prod
\begin{Sb}
 i\\
\reg=0
\end{Sb}
\frac{1}{2}(e^{kx_i}+1)
\prod
\begin{Sb}
 i\\
\reg \ne 0
\end{Sb}
e^{k\reg x_i}
\prod_{\alpha \in\Delta_{+}}
\left\langle \alpha,\phibar\right\rangle^{-2\g}.\label{dform}
\end{align}
%

%%%%%%%%%%%%%%%%%%%%%%%%%%%%%%%%%
%
%
%
The residue formula of the correlation function is given by
\begin{equation}
\left\langle e^{k\omega}\text{Tr}B(\phibar),
\big[{\frak M}_{g}(n,d)\big] \right\rangle =
%\prod_{i=1}^{n-1} 
\underset{\{x_i=0\}}{\text{Res}}\left(\Omega_{g}^{(k)}(n,d)
\text{Tr}B(\phibar)\right).\label{magresidue}
\end{equation}
%%%%%%%%
The order of evaluations of residues above is 
$x_{n-1}<x_{n-2}<\cdots < x_1$.\\
The equivalence  of the multiple infinite sum formula (\ref{sumformula})  
and the residue formula  (\ref{magresidue})  
can be seen using  the localization of infinite  residues sum argument    
as follows.\\
First substituting  the partial fraction expansion  in the residue form 
\begin{align}
\frac{e^{\reg x_i}}{e^{x_i}-1}&=
\sum_{m\in{\Bbb Z}}e^{m\frac{2\pi\sqrt{-1}}{n}di}\dfrac{1}{x_i-2\pi\sqrt{-1}m},
\ \ \reg \ne 0  \label{partialfrac}           \\
\dfrac{1}{2}\frac{e^{x_i}+1}{e^{x_i}-1}&
=\sum_{m\in {\Bbb Z}}\dfrac{1}{x_i-2\pi\sqrt{-1}m},
\end{align}
we can see that the residue of  the form in the right hand side 
of (\ref{magresidue}) evaluated  at any 
dominant integral weight  shifted the weight by $\rho$, 
$\left\{ (l_1,..,l_{n-1})|l_i\geq 1 \right\}$,
coincides with a corresponding summand in (\ref{sumformula})
up to a constant;
\begin{align}
%\prod_{i=1}^{n-1} 
&\underset{\{x_i=2\pi\sqrt{-1}l_i\}}{\text{Res}}\left(\Omega_{g}^{(1)}(n,d)
\text{Tr}B(\phibar)\right)\nonumber \\
&=(-1)^{(n-1)}\frac{1}{n}\times 
 (-1)^{(n-1)d+|\Delta_{+}|\g}n^{g}
\prod_{\alpha \in \Delta_{+}}\left<\alpha,\Phi \right>^{-2\g}
\text{Tr}B(\Phi)
e^{-d\left<\lambda_{1},\Phi \right>}.\label{corresp}
\end{align}
The set of the dominant integral weights shifted by $\rho$
constitute the lattice points set of one of 
the $n!$ Weyl chambers and  (\ref{corresp}) is invariant under 
the Weyl group action. \\
Thus  we can sum over residues over the 
$n!$ Weyl chambers instead of the single chamber as in (\ref{sumformula}).
Let $H_{\alpha}$ be the subset of the weight lattice
perpendicular to the root $\alpha$,\\
$H_{\alpha}=\{(l_i)|l_i\in {\Bbb Z},\langle\Phi,\alpha \rangle=0 \}$.
Then the union of the lattice points of
the $n!$ Weyl chambers coincides with 
the complement in the weight lattice
of the union of the hypersurfaces:
 $L=\overline{\underset{\alpha\in \Delta_{+}}{\bigcup}H_{\alpha}},$ 
and we have
\begin{align}
&\sum_{(l_i)\in L}
%\underset{l_2\ne 0,-l_1}{\sum_{l_2}}\cdots
%\underset{l_3\ne 0,-l_2,-l_1-l_2}{\sum_{l_3}}
%\underset{l_{n-1}\ne 0,-l_{n-2},..,-l_1-l_2-\cdots -l_{n-2}}{\sum_{l_{n-1}}}
\underset{\{x_1=2\pi\sqrt{-1}l_1\} }{\text{Res}}\cdots
\underset{\{x_{n-1}=2\pi\sqrt{-1}l_{n-1}\} }{\text{Res}}
\left(\Omega_{g}^{(1)}(n,d)
\text{Tr}B(\phibar)\right) \nonumber \\
%=&(-1)^{(n-1)}(n-1)!\sum_{l_1=1}^{\infty}
%\cdots \sum_{l_{n-1}=1}^{\infty}(-1)^{(n-1)d+|\Delta_{+}|\g}n^{g}
%\prod_{\alpha \in \Delta_{+}}\left<\alpha,\Phi \right>^{-2\g}
%\text{Tr}B(\Phi)
%e^{-d\left<\lambda_{1},\Phi \right>}\nonumber \\
=&(-1)^{(n-1)}(n-1)!\left\langle e^{\omega}\text{Tr}B(\phibar),
\big[{\frak M}_{g}(n,d)\big] \right\rangle.\label{originalsum}
\end{align}
Now  using the standard residue theorem which  tells that the
total sum of the residues for  one variable 
with the remaining variables fixed is zero \cite{Szenes}
repeatedly from $x_{n-1}$ to $x_1$,
we can reduce the original residues sum (\ref{originalsum}) to  that over
sets  of the form: 
$
\underset{\alpha\in I   }{\bigcap}H_{\alpha} \cap \ %
\overline{\underset{\beta \not\in I  }{\bigcup}H_{\beta}},
$ \\
where  each $I$, beginning   with  $\emptyset$, eventually  becomes
$\Delta_{+}$ in this process.
At last   (\ref{originalsum}) is expressed by the single residue 
evaluated at 
the origin$=\underset{\alpha\in \Delta_{+}}{\bigcap}H_{\alpha}$.
%\begin{equation}
%\underset{\{x_i=0\}}{\text{Res}}\left(\Omega_{g}^{(1)}(n,d)
%\text{Tr}B(\phibar)\right)
%=\frac{(-1)^{(n-1)}}{(n-1)!}
%\sum_{(l_i)\in L}
%\underset{\{x_i=2\pi\sqrt{-1}l_i\} }{\text{Res}}
%\left(\Omega_{g}^{(1)}(n,d)
%\text{Tr}B(\phibar)\right).\label{totalsum}
%\end{equation}
%From (\ref{corresp}) and (\ref{totalsum}) 
Thus the equivalence of  the 
infinite sum formula (\ref{sumformula})
and the residue formula (\ref{magresidue}) 
 follows for $\text{Tr}B(\phibar)$ such that
 (\ref{sumformula}) is convergent.
We also  conjecture that the residue formula (\ref{magresidue})
is valid for arbitrary $\text{Tr}B(\phibar)$.
One evidence for this conjecture is  the fact   
that if we insert  the $\hat{A}$-genus 
${\prod_{\alpha \in\Delta_{+}}
\left(
\frac{\left\langle \alpha/2,\phibar\right\rangle}
{\sinh\left(\left\langle\alpha/2,\phibar\right\rangle\right)}
\right)^{2\g}}$
\cite{Szenes} as a gauge invariant zero form operator in (\ref{magresidue}),
\begin{equation}
\underset{\{x_i=0\}}{\text{Res}}\left(\Omega_{g}^{(k)}(n,d)
\prod_{\alpha \in\Delta_{+}}
\left( \dfrac{\left\langle\alpha/2,\phibar\right\rangle }{
\sinh\left(\left\langle\alpha/2,\phibar\right\rangle\right) }
\right)^{2\g}\right) \label{rational}
\end{equation}
gives the twisted Verlinde dimension${}^{\text{c}}$
 \cite{Verlinde}
 of current algebra of level $ (k-n)$ 
for any $k$ such that
 $k\equiv 0 \mod{\dfrac{n}{(n,d)}}$ in accord with
the prediction of the Riemann-Roch formula, the existence of which is 
also conjectural for the cases of $(n,d)\ne 1$.

%
%
%%%%%%%%%%%%%%%
%%%%%%
\newpage
\subsection{Some examples}
%%%%%%
Here we will give some explicit form of the residue formulas.\\
%%%%%%%%%%%%%%%%%%%%%%%%%%%%%%%%%%%%%%%%%%%%%%%%%%%%%%%%%%
   First for $SU_2$ gauge group, the diagonalization of bosonic ghost becomes
$$
 \phibar =\dfrac{1}{2}
  \begin{pmatrix}
     x_{1} & 0\\
     0 &  -x_{1}
    \end{pmatrix},\ \ {\cal O}_{2}=\dfrac{1}{4}x_1^2
 $$
%%%%%%%%%%%%%%%%%%%%%%%%%%%%%%%%%%%%%%%%%%%%%%%%%%%%%%%%%%%%%
%and $$ is the generator of the Pontjagin class.\\
The residue forms for $d=0,1$ read as
\begin{align}
\Omega^{(k)}_g(2,0)&=
-\left(-2k \right)^{\g}
dx_{1}(k/2)\frac{e^{kx_1}+1}{e^{kx_1}-1}(x_1)^{-2\g}.\\
\Omega^{(k)}_g(2,1)&=
\left(-2k\right)^{\g}
dx_{1}k\frac{e^{kx_1/2}}{e^{kx_1}-1}
(x_1)^{-2\g}.
\end{align}
The correlation functions which have been  completely solved in
\cite{Thadd,Witten4}
can be elegantly expressed  as follows;
\begin{align}
%sazae san
\sum_{\g=0}^{\infty}\lambda^{\g}\left\langle e^{\omega}e^{a{\cal O}_{2}},
\big[{\frak M}_{g}(2,0)\big] \right\rangle &=
-e^{-\frac{1}{2}a\lambda}\sqrt{{\lambda}/{2}}
\cot\left(\sqrt{{\lambda}/{2}}\right)  \\
\sum_{\g=0}^{\infty}\lambda^{\g}\left\langle e^{\omega}e^{a{\cal O}_{2}},
\big[{\frak M}_{g}(2,1)\big] \right\rangle &=
 e^{-\frac{1}{2}a\lambda}\frac{\sqrt{{\lambda}/{2}}}
{\sin\left(\sqrt{{\lambda}/{2}}\right)}.
\end{align}
The generating function of $SU_2$ correlators
of all genera are also considered in \cite{Zagier2}.\\
%%%%%%%%%%%%%%%%%%%%%%%%%%%%%%%%%%%%%%%%%%%%%%%%%%%%%%%%%%%%%%%%%%%%
Next for  $ SU_3 $, the diagonalization of $\phibar$ becomes,     
$$
\phibar =
\dfrac{1}{3}\begin{pmatrix}
2x_{1}+x_{2}& 0 & 0\\
0 & -x_{1}+x_{2}& 0 \\
0 & 0 & -x_{1}-2x_{2}
\end{pmatrix},
$$
$$
    {{\cal O}_{2}}=\dfrac{1}{3}(x_1^2+x_1x_2+x_2^2),\ %
{\cal O}_{3}=\dfrac{1}{54}(2x_1+x_2)(x_1+2x_2)(x_1-x_2).
$$     

%%%%%%%%%%%%
%
%
%
%
The residue forms for  $SU_3$  are given by
\begin{align}
\Omega^{(k)}_g(3,0)&=
\left(-3k^{2}\right)^{\g}
dx_1dx_2(k/2)^2\frac{e^{kx_1}+1}{e^{kx_1}-1}
\frac{e^{kx_2}+1}{e^{kx_2}-1}
(x_1x_2(x_1+x_2))^{-2\g}.\\
\Omega^{(k)}_g(3,1)&=
\left(-3k^{2}\right)^{\g}
dx_{1}dx_{2}k^2
\dfrac{e^{kx_1/3}}{e^{kx_1}-1}
\dfrac{e^{2kx_2/3}}{e^{kx_2}-1}(x_1x_2(x_1+x_2))^{-2\g}.
\end{align}
We give   two simple  examples
 of correlation functions computed by the residue formula; 
\begin{align}
& \left\langle e^{k\omega}e^{a{\cal O}_{2}}e^{b{\cal O}_3},
\big[{\frak M}_{3}(3,0)\big] \right\rangle\nonumber \\
=&\frac{19}{41513472000}k^{16}
-\frac{1}{53222400}ak^{14}
+\frac{1}{2419200}a^2k^{12}
\nonumber \\
-&\left(\frac{1}{120960}a^3+\frac{1}{4354560}b^2\right)k^{10}
+\left(\frac{1}{17280}a^4-\frac{1}{31104}ab^2\right)k^8
\nonumber \\
+&\left(\frac{1}{2592}a^5-\frac{1}{46656}a^2b^2\right)k^6
+\left(\frac{1}{3888}a^6-\frac{7}{34992}a^3b^2
+\frac{107}{7558272}b^4\right)k^4
\end{align}
\begin{align}
& \left\langle e^{3m\omega}e^{a{\cal O}_{2}}e^{b{\cal O}_3},
\big[{\frak M}_{3}(3,1)\big] \right\rangle\nonumber \\
=&\frac{9708939}{512512000}m^{16}-\frac{160911}{1971200}am^{14}
+\frac{1}{896}bm^{13}+\frac{15363}{89600}a^2m^{12}
\nonumber \\
-&\frac{3}{320}abm^{11}+\left(-\frac{1011}{4480}a^3
+\frac{83}{53760}b^2\right)m^{10}
+\frac{21}{640}a^2bm^9\nonumber \\
+&\left(\frac{123}{640}a^4-\frac{5}{384}ab^2\right)m^8
-\left(\frac{5}{96}a^3b-\frac{7}{10368}b^3\right)m^7
+\left(-\frac{3}{32}a^5+\frac{7}{192}a^2b^2\right)m^6
\nonumber \\
-&\left(-\frac{1}{48}a^4b+\frac{19}{2592}ab^3\right)m^5
+\left(\frac{1}{48}a^6-\frac{7}{432}a^3b^2+\frac{107}{93312}b^4\right)m^4
\end{align}
\vspace{0cm}\\
Finally for $SU_4$, the diagonalization of $\phibar$ becomes
$$
\phibar =
\dfrac{1}{4}\begin{pmatrix}
3x_{1}+2x_{2}+x_3 & 0 & 0 & 0\\
0 & -x_1+2x_{2}+x_{3}& 0 & 0\\
0 & 0 & -x_{1}-2x_{2}+x_3 & 0\\
0 & 0 & 0  & -x_1-2x_2-3x_3 
\end{pmatrix}.
$$
In this case we have three different theories with  magnetic flux
units  $d=0,1,2$.
\begin{align}
\Omega^{(k)}_g(4,0)&=
-\left(4k^{3}\right)^{\g}
dx_{1}dx_{2}dx_{3}
(k/2)^3\frac{e^{kx_1}+1}{e^{kx_1}-1}
\frac{e^{kx_2}+1}{e^{kx_2}-1}
\frac{e^{kx_3}+1}{e^{kx_3}-1}
\nonumber \\
&(x_1x_2x_3(x_1+x_2)(x_2+x_3)(x_1+x_2+x_3))^{-2\g}\\
\Omega^{(k)}_g(4,1)&=
\left(4k^{3}\right)^{\g}
dx_{1}dx_{2}dx_{3}k^3
\dfrac{e^{kx_1/4}}{e^{kx_1}-1}
\dfrac{e^{2kx_2/4}}{e^{kx_2}-1}
\dfrac{e^{3kx_3/4}}{e^{kx_3}-1}\nonumber \\
&(x_1x_2x_3(x_1+x_2)(x_2+x_3)(x_1+x_2+x_3))^{-2\g}\\
\Omega^{(k)}_g(4,2)&=
-\left(4k^{3}\right)^{\g}
dx_{1}dx_{2}dx_{3}(k/2)k^2
\dfrac{e^{2kx_1/4}}{e^{kx_1}-1}
\dfrac{e^{kx_2}+1}{e^{kx_2}-1}
\dfrac{e^{2kx_3/4}}{e^{kx_3}-1}\nonumber \\
&(x_1x_2x_3(x_1+x_2)(x_2+x_3)(x_1+x_2+x_3))^{-2\g}
\end{align}

%%%%%%%%%

%%%%%%%%%%%%%%%%%%%%%%%%
\newpage
%%%%%%%%%%%%%%%%%%%%%%%%%%%%%%%%%%%%%%%%%%
\subsection{Bernoulli Expansions}
In principle,  by substituting  in 
 (\ref{magresidue}) the Fourier expansions \cite{Yamaguchi}; 
\begin{align}
\frac{1}{2}\frac{e^{x}+1}{e^{x}-1}
&=\sum_{m=0}^{\infty}B_{2m}\frac{x^{2m}}{(2m)!}
 \\
\dfrac{xe^{\reg x}}{(e^{x}-1)}
&=\sum_{m=0}^{\infty}B_{m}(\theta(d\cdot i/n))
\frac{x^{m}}{m!},\ \ \reg\ne 0
\end{align}
 we {\it can} express any correlation functions by
a finite sum of $(n-1)$-ple products of  Bernoulli polynomials.
% \cite{Yamaguchi}. 
Here we will present the simplest ones.\\
To this end it is convenient to introduce the following notations;
\begin{align*}
b_m(0)&=\frac{B_{m}}{m!}\ \text{for}\  m\ne 1,\ \text{and}\ b_{1}(0)=0, \\
b_m(\theta(d\cdot i/n))&=\frac{B_m(\theta(d\cdot i/n) )}{m!}
\end{align*}
Then the  correlation function of $SU_3$ theory can be written as
a sum of double products of Bernoulli polynomials;
\begin{align}
%\text{Vol}({\frak M}_{g}(3,0))=
&\left\langle e^{\omega}x_1^{a_1}x_2^{a_2},
\big[{\frak M}_{g}(3,d)\big] \right\rangle\nonumber \\
=&(-3)^{\g}\sum_{m_1+m_2=6\g}(-1)^{m_2}
{}_{2\g}\text{H}_{2\g-m_2}b_{m_1-a_1}(\theta(d/3))
b_{m_2-a_2}(\theta(2d/3)).
\end{align}
%
%%%%%%%%%%%%%%%%%%%%%%%%%%%%%%%%%%%%%%%%%%%%%%%%%%%
%%%%%%%%%%%%%%%%%%%%%%%%%%%%%%%%%%55
Similarly the correlation function of $SU_4$ theory is given by
a sum of triple products of Bernoulli polynomials;
\begin{align}
&\left\langle e^{\omega}x_1^{a_1}x_2^{a_2}x_3^{a_3},
\big[{\frak M}_{g}(4,d)\big] \right\rangle\nonumber \\
=&(-1)^{d-1}(4)^{\g}\sum_{m_1+m_2+m_3=12\g}
\sum
\begin{Sb}
l_1,l_2,l_3\geq 0 \\
l_1+l_3=m_1-6\g
\end{Sb}
(-1)^{l_1+l_2+l_3}
{}_{2\g}\text{H}_{l_1}{}_{2\g}\text{H}_{l_2}
{}_{2\g}\text{H}_{l_3}\ {}_{l_3}\text{C}_{2\g-l_2-m_3}\nonumber \\
&b_{m_1-a_1}(\theta(d/4))
b_{m_2-a_2}(\theta(2d/4))b_{m_3-a_3}(\theta(3d/4)).
\end{align}
In this way we can express any correlation function
of $\omega$ and $\phi$ as a finite sum of known rational numbers.
It would be interesting if we understand the relevance of  
the arithmetic properties of Bernoulli numbers \cite{Zagier,Yamaguchi} 
 to two dimensional gauge theories.

%%%%%%%%%%%%%%%%%%%%%%%%%%%%%%%%%%%%%%%%%%%%%%%%%%%%%%%%%%

%%%%%%%%%%%%%%%%%%%%%%%%%%

\newpage 
%%%%%%%%%%%%%%%%%%%%%%%%%%%%%%%%%%%%%%%%%%%%%%%%%%%%%

\newpage
%%%%%%%%%%%%%%%%%%%%%
\section{Deformations by  Two Form Operators}
\subsection{Witten's Formula}
So far we have treated correlation functions which contain 
arbitrary zero operators but
do not contain any two form operator other than  
 the standard symplectic form  $\exp(k\omega)$ .
Now we describe the computation of correlators with arbitrary
two form operator following \cite{Witten4}.
Here again the residue method will turn out to be  useful.\\
Let ${\cal O}$ be a gauge invariant polynomial of 
$\{\phibar^{a}\} $ of the form,
\begin{equation}
{\cal O}={\cal O}_{2} + \sum_{m\geq 3}c_{m}{\cal O}_{m},
\end{equation}
and ${\cal O}^{(2)}$ be the associated two form operator.%\\
\begin{equation}
{\cal O}^{(2)}=-\int_{\Sigma}\left(\frac{1}{2} M_{ab}\psibar^{a}\psibar^{b}
+\frac{\partial{\cal O}}{\partial\phibar^a}\Fbar^a\right),\ \ %
M_{ab}=\frac{\partial^{2}{\cal O}}{\partial\phibar^a\partial\phibar^b}.
\end{equation}
The insertion of $\exp({\cal O}^{(2)})$ in the correlator
corresponds to the deformation of the original Lagrangian by the
two form operator.\\
%We want to compute correlators of the form
%\begin{equation*}
%$\left\langle  \exp({\cal O}^{(2)})\text{Tr}B(\phibar),     
%    \big[{\frak M}_{g}(n,d)\big] \right\rangle $.\\
%%\end{equation*}
By computing the fermion determinant and the Jacobian
of the change of bosonic variables, 
Witten gave the following formula \cite{Witten4}
\begin{equation}
\left\langle  e^{k{\cal O}^{(2)}}\text{Tr}B(\phibar),     
    \big[{\frak M}_{g}(n,d)\big] \right\rangle 
=\left\langle e^{k\omega}
\text{det}M(Q(\phibar))^{\g}
\text{Tr}B(Q(\phibar)),     
    \big[{\frak M}_{g}(n,d)\big] \right\rangle,\label{withtwoform}
\end{equation}
where $Q(\phibar)$ is the power series  defined by the change of 
variables;
\begin{equation}
\widehat{\phi}^a=\widehat{\phi}^a(\phibar)
\equiv\frac{\partial{\cal O}}{\partial\phibar^a},
\ \ \ %
\phibar^a=Q^a(\widehat{\phi}).
\end{equation}
%
%%%%%%%%%%%%%%%%%%%%%%%%%%%%%%%%%%%%%%%%%%%%%%%%%%%%%%%%%%%%%%%%%
\subsection{Inversion of Variables and Residues}
At first sight it might seem necessary to
 convert  the original field variable $\phibar $
into a power series of $\widehat{\phi}$
in order to evaluate the right hand side of (\ref{withtwoform}).
But it is sufficient  to
find   only the inversion of the gauge invariants  $\{{\cal O}_m \}$.
The gauge  invariants constructed by
$\{\widehat{\phi}\}$  are defined by
$
\widehat{{\cal O}}_{m}(\phibar)={\cal O}_{m}(\widehat{\phi}),\ \ 
2\le m \le n.
$
The two sets of Casimir invariants
$\{{\cal O}_{m}\}$ and $\{\widehat{{\cal O}}_{m}\}$ are related by 
certain  polynomial equations:
$
\widehat{{\cal O}}_m=F_m({\cal O}_2,...,{\cal O}_n),
$
and we have  only to convert them  
to evaluate the right hand side of (\ref{withtwoform})
$
{\cal O}_m=G_{m}(\widehat{{\cal O}}_2,...,\widehat{{\cal O}}_n).
$\\
Now the expansion coefficients of  $G_m$ defined by
\begin{equation}
G_m(p_2,...,p_n)=\sum_{l_2,...,l_n\geq 0}G_{m}(l_2,..,l_n)p_2^{l_2}\cdots
p_n^{l_n},
\end{equation}
can be obtained using the Cauchy formula \cite{MoFe};
\begin{align}
G_{m}(l_2,..,l_n)&=
\underset{\{p_i=0\}}{\text{Res}}
\left(G_{m}(p)/(p_2^{l_2+1}\cdots p_n^{l_n+1})dp_2\cdots dp_n\right)
\nonumber \\
&=\underset{\{q_i=0\}}{\text{Res}}\left(q_m
\text{det}\left(\frac{\partial F_i}{\partial q_j}\right)
/(F_2^{l_2+1}(q)\cdots F_{n}^{l_n+1}(q))
dq_2\cdots dq_n\right).
\end{align}
Thus  we get at least formally 
the following residue formula
for the power series $G_m$ of $p_i$  
\begin{equation}
G_m(p_2,...,p_n)=\underset{\{q_i=0\}}{\text{Res}}\left(
q_m \text{det}\left(\frac{\partial F_i}{\partial q_j}\right)
\prod_{i=2}^{n}\frac{dq_i}{(F_i(q)-p_i)}\right).\label{mofe}
\end{equation}
\subsection{Diagonalization}
To get the explicit polynomial relations $\{F_m\}$ of the previous subsection
between the old and new Casimir invariants,
it suffices to know only the change of variables for the diagonalization
of the fields  $\phibar$ because of  gauge invariance;
\begin{align}
\widehat{\phi}&=\sum_{i=1}^{n-1}\widehat{x}_iH_i
=\text{diag}(\widehat{z}_1,..,\widehat{z}_n),\ \ % 
\widehat{\cal O}_l=\sum_{i=1}^{n}\frac{1}{l!}(\widehat{z}_i)^{l},\nonumber \\ 
\widehat{x}_i &=y_i(x)=C_{ij}\frac{\partial {\cal O}}{\partial x_j}
=(z_i-z_{i+1})+\sum_{m\geq 2}
\frac{c_{m+1}}{m!}(z_i^{m}-z_{i+1}^{m}),\label{change}\\
\widehat{z}_i&=z_i+\sum_{m\geq 2}\frac{c_{m+1}}{m!}z_i^{m}-\frac{1}{n}
\sum_{m\geq 2}c_{m+1}{\cal O}_m\nonumber
\end{align}
We  also have the following determinant formula by the diagonalization;
\begin{equation}
\text{det}M(\phibar)=n\text{det}
\left(\frac{\partial^2 {\cal O}}{\partial x_i \partial x_j}\right)
\prod_{\alpha\in \Delta_{+}}
\left(\frac{\langle \alpha, \widehat{\phi}\rangle}
{\langle \alpha, \phibar \rangle}\right)^2.\label{detform}
\end{equation}
Now we can compute any correlation functions
(\ref{withtwoform}) using (\ref{mofe}),(\ref{change}) and (\ref{detform}).\\
The consistency of our formalism may be checked by considering
 deformations of $SU_2$ theory because in $SU_2$ theory any observable
can be expressed by  $\omega,{\cal O}_{2}$ and $V_2(a)$.\\
For example the two form observable
associated with ${\cal O}={\cal O}_2-(2l-1)!b_l{\cal O}_{2l}$ 
can be written by the observables associated with the Casimir invariant 
of seccond degree as
\begin{equation}
{\cal O}^{(2)}=
\omega-b_l\left({\cal O}_{2}^{l-1}\omega-(l-1){\cal O}_{2}^{l-2}\sum_{a=1}^{g}
V_2(a)V_2(a+g)\right) \label{reducedtwo}.
\end{equation}
It can be seen that the use of 
the Witten's formula and the direct substitution
of (\ref{reducedtwo}) in the left hand side of (\ref{withtwoform})
give the same answer.\\ 
Next take, for example, the $SU_3$ theory with the two form operator 
associated with
${\cal O}={\cal O}_2-6m{\cal O}_3.$
The polynomial relation between old and new Casimir invariants is given by 
%\begin{align}
\begin{equation}
\begin{cases}
\widehat{{\cal O}}_2 &={\cal O}_2-18m{\cal O}_3+3m^2{\cal O}_2^2 \\
\widehat{{\cal O}}_3 &={\cal O}_3-m{\cal O}_2^2+9m^2{\cal O}_2{\cal O}_3
+m^3({\cal O}_2^3-54{\cal O}_3^2).\label{su3defo}
\end{cases}
%\end{align}
\end{equation}
>From the residue formula (\ref{mofe}), we get the inversion 
of the polynomial relation (\ref{su3defo}) as follows;
%\begin{align}
\begin{equation}
\begin{cases}
{\cal O}_2 &=
\widehat{{\cal O}}_2
+18m\widehat{{\cal O}}_3
+15m^2\widehat{{\cal O}}_2^2
+378m^3\widehat{{\cal O}}_3\widehat{{\cal O}}_2
+m^4(2916\widehat{{\cal O}}_3^2+270\widehat{{\cal O}}_2^3)+\cdots, \\
{\cal O}_3 &=
\widehat{{\cal O}}_3
+m\widehat{{\cal O}}_2^2
+27m^2\widehat{{\cal O}}_2\widehat{{\cal O}}_3 
+m^3(216\widehat{{\cal O}}_3^2+20\widehat{{\cal O}}_2^3)+
810m^4\widehat{{\cal O}}_2^2\widehat{{\cal O}}_3+\cdots.
\end{cases}
\end{equation}
%\end{align}
The determinant that appears in Witten's formula is given by 
\begin{equation}
\text{det}M=(1-12m^2{\cal O}_2)(1-9m^2{\cal O}_2+54m^3{\cal O}_3)^2.
\end{equation}
Then by the formula (\ref{withtwoform}) we get the results 
\begin{align}
 \left\langle e^{3a{\cal O}^{(2)}},
\big[{\frak M}_{2}(3,1)\big] \right\rangle
=&\frac{477}{2240}a^8+\frac{189}{8}a^6m^2-27a^5m^3
-\frac{405}{4}a^4m^4 \\
 \left\langle e^{3a{\cal O}^{(2)}},
\big[{\frak M}_{3}(3,1)\big] \right\rangle
=&\frac{9708939}{512512000}a^{16}+\frac{482733}{98560}a^{14}m^2
-\frac{27}{28}a^{13}m^3
+\frac{1244403}{4480}a^{12}m^4\nonumber \\
-&243a^{11}m^5-\frac{38151}{16}a^{10}m^6
-\frac{423549}{80}a^8m^8.
\end{align}
\subsection{Generalized Residue  Formula }
One method to compute the correlators with a general two form operator 
was to expand the Casimir invariants
$\{{\cal O}_m\}$ into the series of $\{\widehat{{\cal O}}_m\}$
perturbatively and substitute them in the right hand side of
(\ref{withtwoform}).
Here we give another method in the form of residue formula.
First   we use the residue formula of the previous section
in the right hand side of (\ref{withtwoform}) to obtain 
\begin{equation}
\left\langle e^{k{\cal O}^{(2)}}\text{Tr}B(\phibar),
\big[{\frak M}_{g}(n,d)\big] \right\rangle =
\underset{\{y_i=0\}}{\text{Res}}\left(\Omega_{g}^{(k)}(n,d)
\text{Tr}B(Q(y))\text{det}M^{\g}(Q(y))\right).
\end{equation} 
Then  if we change the integration variables 
from  $y_i=\widehat{x}_i$ to $Q_i(y)=x_i$, 
we obtain the generalized residue formula; 
\begin{equation}
\left\langle e^{k{\cal O}^{(2)}}\text{Tr}B(\phibar),
\big[{\frak M}_{g}(n,d)\big] \right\rangle =
\underset{\{x_i=0\}}{\text{Res}}
\left(\Omega_{g}^{(k)}(n,d;{\cal O})
\text{Tr}B(x)
%\text{det}\left(\frac{\partial y_j}{\partial x_i}\right)^g(x)
\right),\label{genres}
\end{equation}
where 
\begin{align}
&\Omega_{g}^{(k)}(n,d;{\cal O}) =
(-1)^{(n-1)(d-1)+|\Delta_{+}|\g}
\left(nk^{n-1}\right)^{\g}
\prod_{i=1}^{n-1} dx_i
\dfrac{k}{(e^{ky_i}-1)}\nonumber \\
& J^g \prod
\begin{Sb}
 i\\
\reg=0
\end{Sb}
\frac{1}{2}(e^{ky_i}+1)
\prod
\begin{Sb}
 i\\
\reg \ne 0
\end{Sb}
e^{k\reg y_i}
\prod_{\alpha \in\Delta_{+}}
\left\langle \alpha,\phibar\right\rangle^{-2\g}\nonumber \\
=& J^g \prod_{i=1}^{n-1}
\left(\frac{e^{kx_i}-1}{e^{ky_i}-1}\right)
\prod
\begin{Sb}
 i\\
\reg=0
\end{Sb}
\left(\frac{e^{ky_i}+1}{e^{kx_i}+1}\right)
\prod
\begin{Sb}
 i\\
\reg \ne 0
\end{Sb}
e^{k\reg (y_i-x_i)}
\Omega_{g}^{(k)}(n,d)
\end{align}
and the Jacobian is 
\begin{equation}
J=\text{det}\left(\frac{\partial y_j}{\partial x_i}\right)
=n\text{det}\left(\frac{\partial^2 {\cal O}}{\partial x_i \partial x_j}\right).
\end{equation}
Thus we can say that the deformation of topological Yang-Mills theory
by a two form operator is equivalent to the insertion of a certain 
zero form operator.
Note that in this section the reduction of the gauge group to the 
abelian subgroup \cite{BlaTho2,BlaTho2.5} was the powerful tool 
to compute explicitly the 
various physical quantities.
%

%%%%%%%%%%%%%%%%%%%%%%%%%%%%%%%%%%%%%%%%%%%%%%%%%%%%%%%%%%%
%
\section{Recursion Relation}
\subsection{Wick Contraction of One Form Operators}
Here as an application of the 
residue formula described above 
 we will consider correlation functions containing  one form observables.
It is not difficult to compute them because  we can use
the physical Yang-Mills 
Lagrangian  (\ref{PhysYangMills})   to integrate out
all one form observables
 in correlators of the topological theory \cite{Witten4}
owing  to the  physical/topological 
Yang-Mills correspondence.
Indeed  using the gauge invariance and the diagonalization
we get  the following contraction formula in the presence of 
the two form operator  $\exp(k{\cal O}^{(2)})$;
\begin{equation}
\left\langle V_{m}(a)V_{l}(b+g)\right\rangle=
-\dfrac{1}{k}\delta_{ab}\sum_{ij}H^{-1}_{ij}
\frac{\partial{\cal O}_{m}}{\partial x_i}
\frac{\partial{\cal O}_{l}}{\partial x_j}, 
\label{Wick}
\end{equation}
where $H_{ij}$ is the Hessian 
$
H_{ij}=\dfrac{\partial^2 {\cal O}}{\partial x_i\partial x_j}$.
%
%
%
%%%%%%%%%%%%%%%%%%%%%%%%%%%%%%%%%%%%%%%%%%%%%%%%%%%%%%%%%%%%%%%%%%%%%%%%%%
\subsection{ Handle Contracting  Operator; $SU_2$}
In \cite{Thadd} Thaddeus rigorously proved 
that as a cohomology class $V_{2}(a)$ is the Poincare dual of the 
subspace $N_g(a)$ of ${\frak M}_{g}(2,1)$ where the holonomy around 
the cycle $C_a$ is trivial.\\
The physical meaning of it
is that  $V_{2}(a)$ has only the effect of reducing the 
path integral to the flat gauge fields 
that  have the trivial holonomy around $C_a$.
Thus   $V_{2}(a)$  may    be  regarded as a operator
 which contracts the cycle $C_a$.
The cup product of them  $V_{2}(a)V_{2}(a+g)$ 
is the Poincare dual to 
$N_g(a)\cap N_{g}(a+g)$  where the holonomies around the both
cycles $C_a$ and $C_{a+g}$ are trivial.\\
Thus we call here $ H_2(a)\equiv V_{2}(a)V_{2}(a+g)$ the operator 
that contracts the $a$-th handle.\\
Noting that $N_g(a)\cap N_{g}(a+g)$
is {\it diffeomorphic} to ${\frak M}_{g-1}(2,1)$,
	 we have the  relation between correlators of genus $g$ and $g-1$,  
\begin{equation}
\left\langle V_{2}(a)V_{2}(a+g)(\cdots),
\left[{\frak M}_{g}(2,1)\right]\right\rangle
=\left\langle (\cdots),
\left[{\frak M}_{g-1}(2,1)\right]\right\rangle,\label{su2relation}
\end{equation}
where $(\cdots)$ means any operators. 
$H_2(a)$ is the {\it inverse} of the handle operator
in the ordinary topological field theories in two dimensions.\\
The physical derivation of this effect  \cite{Witten4} is as follows.
Consider a  correlation function of $SU_2$ theory ; 
$
\left\langle V_{2}(a)V_{2}(a+g)\text{Tr}B(\phibar)e^{k\omega},
\left[{\frak M}_{g}(2,d)\right]\right\rangle.
$
We can easily integrate out the one form operators
if we use  the physical Yang-Mills Lagrangian (\ref{PhysYangMills})
which produces the trivial propagator for the fermions and then return to
the topological Lagrangian to get 
$\left\langle V_{2}(a)V_{m}(a+g)\right\rangle=-\dfrac{2}{k}{\cal O}_{2} $.\\
Then according to the residue formula (\ref{magresidue})
 it is clear that the
insertion of \\
$-\dfrac{2}{k}{\cal O}_{2}=-\dfrac{x_1^2}{2k}$
   reduce the genus of the surface by one.\\
Thus we have found the formula which is equivalent to  (\ref{su2relation}),
\begin{align}
&\left\langle V_{2}(a)V_{2}(a+g)\text{Tr}B(\phibar)e^{k\omega},
\left[{\frak M}_{g}(2,d)\right]\right\rangle\nonumber \\
=&\left\langle 
-\dfrac{2}{k}{\cal O}_{2}\text{Tr}B(\phibar)e^{k\omega},
\left[{\frak M}_{g}(2,d)\right]\right\rangle 
=\left\langle \text{Tr}B(\phibar)e^{k\omega},
\left[{\frak M}_{g-1}(2,d)\right]\right\rangle
\end{align}
\subsection{ Handle Contracting Operator; Generalization to $SU_n$}
The identification of $V_2(a)$ with the $a$-th cycle contracting
operator of $SU_2,\ d=1$ theory was possible \cite{Thadd} because
$V_2(a)$ is the only observable that satisfies the two requirements:\\
(1) It should have the ghost number $(n^2-1)=3.$\\
(2) It must be fixed by the modular transformations that fix  $C_a$.\\
It seems impossible to extend this pure
topological method to higher rank gauge groups $ SU_n,\ n>3$.
%   
%The operator that contracts the  of $SU_2$ theory 
Nevertheless 
 we can identify even  for  higher rank $SU_n$ theories
  the operator which contracts the $a$-th cycle by using 
the generalized residue formula (\ref{genres})
and the contraction formula of fermions (\ref{Wick}).
We claim that  for general $SU_n$  
 the operator that contracts the $a$-th handle  is the following;
\begin{equation}
H_n(a)=\prod_{l=1}^{n-1}l! \ V_{2}(a)\cdots V_{n-1}(a)
\prod_{l=1}^{n-1}l! \ V_{2}(a+g)\cdots V_{n-1}(a+g).
\end{equation}
Indeed in the presence of the general two form operator 
$\exp(k{\cal O}^{(2)})$, the  Wick contraction
of fermions in  the physical  Yang-Mills theory 
(\ref{PhysYangMills}) gives, 
\begin{align}
&\left\langle H_{n}(a)\right\rangle 
=\left(\prod_{l=1}^{n-1}l!\right)^2
\text{det}\left(\langle V_n(a)V_m(a+g) \rangle\right)\nonumber \\
=&(-1)^{\frac{1}{2}(n-2)(n-1)}\left(\prod_{l=1}^{n-1}l!\right)^2
\left(\frac{-1}{k}\right)^{n-1}\text{det}
\left(\frac{\partial{\cal O}_{m}}{\partial x_i}H^{-1}_{ij}
\frac{\partial{\cal O}_{l}}{\partial x_j}\right)\nonumber \\
=&(-1)^{\frac{1}{2}n(n-1)}
\frac{1}{k^{n-1}}\left(\prod_{l=1}^{n-1}l!\right)^2
\text{det}H^{-1}
\text{det}\left(\frac{\partial{\cal O}_{m}}{\partial x_i}\right)^2.
\end{align}
Furthermore  we can compute the determinants  above as
\begin{align*}
\text{det}\left(\frac{\partial{\cal O}_{m}}{\partial x_i}\right)
=&\text{det}\left(\sum_{j=1}^{n}\frac{\partial z_j}{\partial x_i}
\frac{\partial{\cal O}_{m}}{\partial z_j}\right)
=\underset{1\le i,j\le n-1}{\text{det}}
\left(\frac{\partial z_j}{\partial x_i}\right)
\underset{1\le j,m \le n-1}{\text{det}}
\left(\frac{(z_j^{m}-z_n^{m})}{m!} \right)\\
=&\frac{1}{n}(-1)^{\frac{1}{2}(n-1)(n-2)}
\left(\prod_{l=1}^{n-1}l!\right)^{-1}\prod_{i<j}(z_i-z_j),\\
\text{det}H=&\frac{1}{n}\text{det}
\left(\frac{\partial y_j}{\partial x_i}\right).
\end{align*}
Finally we get the Wick contraction formula
\begin{equation}
\left\langle H_{n}(a)\right\rangle=(-1)^{|\Delta_{+}|}
\text{det}\left(\frac{\partial y_j}{\partial x_i}\right)^{-1}
\frac{1}{nk^{n-1}}\prod_{\alpha\in \Delta_{+}}\langle \alpha,\phibar \rangle^2.
\end{equation}
We see then  according to the generalized residue formula (\ref{genres}),
the insertion of $H_n(a)$ has the effect of 
reducing the genus of the surface by one.
Thus we have the recursion relation
\begin{equation}
\left\langle H_{n}(a)\text{Tr}B(\phibar)e^{k{\cal O}^{(2)}},
\left[{\frak M}_{g}(n,d)\right]\right\rangle
=\left\langle \text{Tr}B(\phibar)e^{k{\cal O}^{(2)}},
\left[{\frak M}_{g-1}(n,d)\right]\right\rangle.
\end{equation}
It is clear that the above equation is  also valid 
even when we insert  any one form operators.\\
Here we describe explicitly the Wick contraction 
for the  case  of   $SU_4$  and ${\cal O}={\cal O}_2$.\\
The handle contracting operator for  $SU_4$ is 
\begin{equation*}
H_{4}(a)=2\cdot 3! \cdot V_{2}(a)V_{3}(a)V_{4}(a)\cdot
2\cdot 3! \cdot V_{2}(a+g)V_{3}(a+g)V_{4}(a+g),
\end{equation*}
and  the Wick contraction of this  becomes
\begin{align}
\left\langle H_{4}(a)\right\rangle&=
-\frac{(2!3!)^2}{k^3}\begin{vmatrix}
-2{\cal O}_2 &  -3{\cal O}_3 & -4{\cal O}_4\\
-3{\cal O}_3 & 1/4{\cal O}_2^2-6{\cal O}_4 & -7/12{\cal O}_2{\cal O}_3\\
-4{\cal O}_4 & -7/12{\cal O}_2{\cal O}_3 & 1/36{\cal O}_2^3-1/12{\cal O}_3^2
+{\cal O}_2{\cal O}_4
\end{vmatrix}\nonumber \\
&=\frac{144}{k^3}\left(\frac{1}{72}{\cal O}_{2}^6
-\frac{5}{6}{\cal O}_{2}^4{\cal O}_{4}
-\frac{17}{36}{\cal O}_{2}^3{\cal O}_{3}^2
+16{\cal O}_{2}^2{\cal O}_{4}^2+6{\cal O}_{2}{\cal O}_{3}^2{\cal O}_{4}
-96{\cal O}_{4}^3-\frac{3}{4}{\cal O}_{3}^4\right)\nonumber \\
&=\frac{1}{4k^3}\left( x_1x_2x_3(x_1+x_2)(x_2+x_3)(x_1+x_2+x_3) \right)^2.
\end{align}
%%%%%%%%%%%%%%%%%%%%%%%%
%\newpage
\section{Footnotes}
\begin{enumerate}
\renewcommand{\labelenumi}{(\alph{enumi})}%
\item Magnetic flux here represents an element of 
 $H^2(\Sigma_{g},{\Bbb Z}_n)\cong
{\Bbb Z}_n$ which classifies the topology of
$SU_n/{\Bbb Z}_n$ bundles on $\Sigma_{g}$.
\item The ample generator of the Picard group 
$\cong{\Bbb Z}$ is $\dfrac{n}{(n,d)}\omega$,
 and  the first  Chern class  is $ 2n\omega$ in any case.
%\item In particular the volume of the gauge group
% in Witten's formula  is \\
%$
%\text{Vol}(SU_n)=\dfrac{n^{\frac{1}{2}}(2\pi)^{\frac{1}{2}(n-1)(n+2)}}%
%{\prod_{i=1}^{n-1}i!}.
%$
\item If  $k \not\equiv 0 \mod{\dfrac{n}{(n,d)}},$ then
the twisted Verlinde dimension is precisely  zero, while
the residue (\ref{rational}) gives a rational number.
\end{enumerate}
 %%%%%%%%%%%%%%%%%%%
%%%%%%%%%%%%%%%%%%%%%%%%%%%%%%%%%%%%%%%%%%%%%%%%%%%%%%%%
%%%%%%%%%%%%%%%%%%%%%%%%
\newpage
%%%%%%%%%%%%%%%%%%%%%%%%%%%%%%%%%%%%%%%%%%%%%%%%%%%%%%%%%%%%

\end{document}